\documentclass[journal]{IEEEtran}
\usepackage{amsmath,amsfonts}
\usepackage{array}
\usepackage[caption=false,font=normalsize,labelfont=sf,textfont=sf]{subfig}
\usepackage{textcomp}
\usepackage{stfloats}
\usepackage{url}
\usepackage{verbatim}
\usepackage{graphicx}
\usepackage{cite}
\usepackage{amsmath,amssymb,amsfonts}
\usepackage{mathrsfs}
\usepackage{setspace,booktabs}
\usepackage{mathalpha}
\usepackage{mathtools,fancyhdr,nccmath}
\usepackage{nicematrix}
\usepackage{arydshln}
\usepackage{algorithm,algpseudocode,float}
\makeatletter
   \let\NAT@parse\undefined
   \makeatother
\usepackage{hyperref}

\hypersetup{hidelinks}
\hypersetup{
    colorlinks,
    linkcolor={blue},
    citecolor={blue},
    urlcolor={blue!80!black},
}

\newtheorem{thm}{Theorem}
\newtheorem{lem}[thm]{Lemma}

\newtheorem{rem}{Remark}
\newtheorem{prob}{Problem}
\def\BibTeX{{\rm B\kern-.05em{\sc i\kern-.025em b}\kern-.08em
    T\kern-.1667em\lower.7ex\hbox{E}\kern-.125emX}}
\usepackage{balance}

\begin{document}
\title{Coordinated Control of Autonomous Vehicles for Traffic Density Reduction at a Signalized Junction: An MPC Approach}
\author{Rudra Sen, \IEEEmembership{Student Member, IEEE}, and Subashish Datta, \IEEEmembership{Member, IEEE}
\thanks{This paragraph of the first footnote will contain the date you submitted your paper for review. This work is supported by the Prime Minister's Research Fellowship scheme under the Ministry of Education, Government of India.}
\thanks{Rudra Sen and Subashish Datta are with the Department of Electrical Engineering, Indian Institute of Technology Delhi, New Delhi, Delhi 110016, India (e-mail: rudra.sen@ee.iitd.ac.in; subashish@ee.iitd.ac.in).}
}

\markboth{Journal of \LaTeX\ Class Files,~Vol.~18, No.~9, September~2020}%
{How to Use the IEEEtran \LaTeX \ Templates}

\maketitle

\begin{abstract}
    The effective and safe management of traffic is a key issue due to the rapid advancement of the urban transportation system. Connected autonomous vehicles (CAVs) possess the capability to connect with each other and adjacent infrastructure, presenting novel opportunities for enhancing traffic flow and coordination. This work proposes a dual-mode model predictive control (MPC) architecture that tackles two interrelated issues: mitigating traffic density at signalized junctions and facilitating seamless, cooperative lane changes in high-density traffic conditions. The objective of this work is to facilitate responsive decision-making for CAVs, thereby enhancing the efficiency and safety of urban mobility. Moreover, we ensure recursive feasibility and convergence of the proposed MPC scheme by the integration of an online-calculated maximal control invariant terminal set. Finally, the efficacy of the proposed approach is validated through numerical simulation.
\end{abstract}


\section{Introduction}
\label{sec:introduction}
\IEEEPARstart{D}{ue} to both the expanding population and the increased need for transportation, traffic congestion has become a significant problem in urban areas. It causes a number of social and environmental issues, such as a rise in traffic accidents, detrimental economic impacts, and elevated greenhouse gas emissions. Currently, in traffic transportation, the most common way to manage traffic flow through a junction is via traffic signals. There are now new approaches that can be used to regulate traffic lights due to recent technological advancements that take advantage of the capacity to collect traffic data in real-time \cite{Z_Liu_2007}. However, the issue of road congestion still persists. The emergence of autonomous vehicles (AVs) ensures smoother traffic flow at various road segments. The collective term of AVs is called \emph{connected autonomous vehicles} (CAVs), as they are connected with other AVs and roadside infrastructure on the road through vehicle-to-vehicle (V2V) and vehicle-to-infrastructure (V2I) communication, respectively. The introduction of CAVs in the transportation systems allows vehicles to make better operation decisions, which leads to a significant reduction in energy consumption and travel delays, and also has the potential to improve safety \cite{S_Melo_2017},\cite{I_Klein_2016}. 

Traffic congestion is the prime issue as the junctions are the converging points of traffic and flow in different directions. In a traffic scenario, aggressive drivers usually accelerate their vehicles to reduce travel time. Such behaviour in congested traffic generates density waves that travel backward with increasing intensity, resulting in a rise in vehicle accumulation near a signal post. This causes one or more queued vehicles to be unable to cross the junction within the specified time span, which increases the wait time of the vehicles, causes travel wariness, and hampers the constant flow of traffic. A centralized coordination structure based on an intersection manager is first presented by Dresner and Stone \cite{K_Dresner_2008}  to effectively schedule autonomous vehicles and mitigate traffic congestion at unsignalized crossings. There is extensive research focusing on the scheduling strategy of CAVs in signal-free junctions. A centralized control strategy in \cite{J_Lee_2012} is designed considering a signal-free junction to minimize the possible overlaps of vehicle trajectories from all directions at the junction. In \cite{H_Xu_2020}, the authors suggested a heuristic search strategy for CAVs to pass through the unsignalized junction, though the given strategy provides sub-optimal results. Liu \emph{et al.} \cite{D_Liu_2025} proposed an adaptive control protocol for a longitudinal vehicular platoon with partial state-feedback capabilities. Bichiou and Rakha \cite{Y_Bichiou_2019} provided a centralized control method to minimize traveling time using Pontryagin's minimum principle. Pan \emph{et al.}\cite{X_Pan_2022} presented a centralized hierarchical CAV coordination technique to minimize trip time and energy consumption. Numerous studies on the coordination of CAVs in signal-free junctions \cite{A_Malikopoulos_2018, I_Mahbub_2019, Y_Zhang_2019, B_Chalaki_2021, B_Chalaki_2021a} have also been documented in the literature employing decentralized optimal control methodologies to derive closed-form solutions. Similarly, to lessen traffic congestion at signalized junctions, a number of control strategies have been proposed, including the multi-intersections-based fuel efficient predictive control method \cite{H_Dong_2022}, model predictive control (MPC) framework \cite{B_Asadi_2010, A_Kamal_2014, J_Han_2018, T_Baby_2022, R_Sen_2024}, optimal control framework \cite{X_Meng_2018}.

Another significant concern in heavy traffic scenarios is the lane change maneuver. This can be intricate, especially in dense traffic situations, since drivers need to assess safety parameters concurrently with the neighboring vehicles. Over the past decades, research on AVs and advanced driver assistance systems (ADAS) has made remarkable advances \cite{VK_Kukkala_2018, L_Masello_2022}.
Methods such as a fuzzy logic-based method \cite{J_Naranjo_2008}, learning based methods \cite{S_Bae_2022, D_Li_2023}, a utility-based method \cite{S_Zeinali_2024}, MPC framework \cite{A_Falsone_2020, A_Falsone_2022, S_Bae_2022, V_Bhattacharyya_2023, T_Skibik_2023, HD_Nguyen_2024, F_Laneve_2024} have found reasonable success in the literature for the lane changing maneuver. Authors of \cite{A_Falsone_2020} propose a lane changing maneuver for an AV in a platoon, wherein they decouple the online computation of the coordination phase from the offline computation of the lane change reference profile for the merging phase to enable MPC-based maneuvering. Subsequently, Falsone \emph{et al.} \cite{A_Falsone_2022} further advanced the coordination phase by computing offline optimal coordinated maneuver using explicit MPC. Bae \emph{et al.} \cite{S_Bae_2022} presented an MPC framework for safe lane change, integrated with recurrent neural network (RNN) and social generative adversarial networks (SGAN) to predict the future position of neighbouring vehicles. Nguyen \emph{et al.} \cite{HD_Nguyen_2024} suggested a robust MPC-based approach with an ellipsoidal set as a terminal constraint for lane changing. A parametric MPC approach to address the lane change maneuver is presented in \cite{F_Laneve_2024}.

In the aforementioned studies, researchers examine the issues of congestion and lane changing of CAVs independently. Furthermore, in the MPC methodologies addressing these challenges, numerous studies have either neglected the feasibility and convergence concerns of MPC or have utilized an ellipsoidal set as a terminal constraint, resulting in sub-optimal solutions. In this paper, we extend the work of \cite{R_Sen_2024}, presented in IEEE CDC, in two ways. First, we propose an algorithm to mitigate the congestion or over-queuing problem during the red signal phase near the junction point. Here, we generate a reference longitudinal velocity for each AV based on the signal information received from the signalized junction. We subsequently formulate the constrained optimization problem to track the reference velocity in the MPC framework. Second, we propose an algorithm to address the safe and cooperative lane change maneuver of AVs in dense traffic situations based on information received from neighbouring vehicles within the control zone. Furthermore, to ensure the feasibility and stability of the MPC approach, we computed an online control-invariant terminal set for each AV at every time instant.

\subsection{Notations} $x^i_{p|k}$ indicates the predicted value of state $x$ of $i^{th}$ AV at time $k+p$ with the information from time $k$. The superscript $(\cdot)^j$ refers to the information of $j^{th}$ AV, and the superscript $(\cdot)^\star$ on terms denotes the optimal value obtained by solving the optimization problem. $\Vert\cdot\Vert$ indicates $2$-norm. For any scalar value $c$, $\overline{c}$ and $\underline{c}$ denote the maximum and minimum value of $c$, respectively. $\text{conv}(V)$ refers to convex hull of all the elements of $V = \{V_1, V_2,\ldots,V_n\}$. Consider $\mathcal{S}$ and $\mathcal{T}$ are the two polyhedral set, the Minkowski sum $\mathcal{S} \oplus \mathcal{T} \triangleq \{s+t~|~s\in\mathcal{S},~t\in\mathcal{T}\}$. Consider $\mathcal{P} = \{x\in\mathbb{R}^n~|~Mx\leq b\}$ is a a polyhedral set with $M\in\mathbb{R}^{n_P\times n}$ and $b\in\mathbb{R}^{n_p}$, and $R\in\mathbb{R}^{n\times n}$ is any matrix. Then $\mathcal{P}\circ R = \{x\in\mathbb{R}^n~|~MRx\leq b\}$ and $R\circ\mathcal{P} = \text{conv}(RV_1, RV_2,\ldots,RV_q)$, where $\mathcal{P} = \text{conv}(V_1, V_2,\ldots, V_q)$ is the vertex representation of $\mathcal{P}$ and $V_1, V_2, \ldots V_q$ are the vertices of set $\mathcal{P}$. Any matrix $S\succ 0~(S\succeq 0, S\prec 0, S\preceq 0)$ denotes a symmetric positive definite (positive semi-definite, negative definite, negative semi-definite) matrix, respectively. $I_q$ is the identity matrix of size $q\times q$. $\Vert r\Vert_S = r^\top Sr$ is weighted $2$-norm of a vector $r$ with weight matrix $S$. $\mathbb{I}_c^d$ denotes the set of integer in the interval $[c, d]$.

\section{Vehicle Model and Problem Formulation}
\label{sec:problem_formulation}
We consider a signalized junction and also assume an autonomous traffic environment where AVs enter the \emph{control zone} at different time instants. Vehicles can change lanes inside that zone. We define the \emph{control zone} as the area where AVs can communicate with the roadside infrastructure through V2I communication, which is typically considered to be a radius of a few hundred meters. A coordinator allocates a unique ID to each AV as they enter the \emph{control zone} at different time instants. However, if two or more AVs enter the zone at the same time, then the coordinator assigns numbers to the AVs randomly. It also removes the AVs from the list upon their departure from the control zone.

In this section, we first briefly describe the dynamics considered for an AV, and thereafter, we formulate the problem.

\subsection{Vehicle Model}
\label{subsec:vehicle_model}
\begin{figure}[thbp]
      \centering
      \includegraphics[height = 4.0cm, width = 7.0cm]{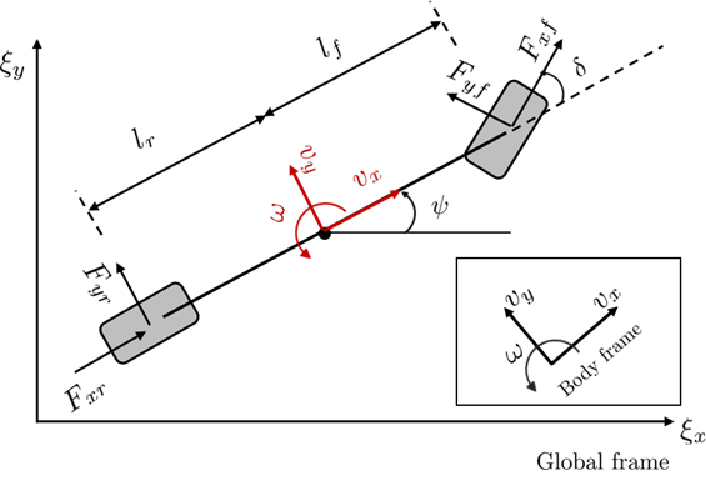}
      \caption{Dynamic Model of AV}
      \label{fig:dynamics_vehicle}
   \end{figure}

We consider a dynamic model of an AV for the $i^{th}$ AV as shown in Fig.~(\ref{fig:dynamics_vehicle}), which describes vehicle motion on the ground while maintaining accuracy and reducing model complexity. The dynamic model in Fig.~(\ref{fig:dynamics_vehicle}) considered only the vehicle's longitudinal, lateral, and yaw motions. 

The dynamics of $i^{th}$ AV can be described as \cite{R_Rajamani_2011, J_Wang_2017}
    \begin{subequations}
        \label{eq:dynamic_model_NL}
        \begin{equation}
            \label{eq:dyn_long_lat_position}
            \Dot{\xi}^i = \begin{bmatrix}\Dot{\xi}_x^i\\\Dot{\xi}_y^i\end{bmatrix} = \begin{bmatrix}\cos{\psi^i}&-\sin{\psi^i}\\ \sin{\psi^i}&\cos{\psi^i} \end{bmatrix} \begin{bmatrix}v_x^i\\v_y^i\end{bmatrix}
        \end{equation}
        \begin{equation}
            \label{eq:dyn_long_velocity}
            \Dot{v}^i_x = v^i_y\omega^i + a^i_x
        \end{equation}
        \begin{equation}
            \label{eq:dyn_lat_velocity}
            \Dot{v}^i_y = -v^i_x\omega^i + \dfrac{F^i_{yf}+F^i_{yr}}{m^i}
        \end{equation}
        \begin{equation}
            \label{eq:dyn_yawAngle}
            \Dot{\psi}^i = \omega^i 
        \end{equation}
        \begin{equation}
            \label{eq:dyn_angularVel}
            \Dot{\omega}^i =  \dfrac{l^i_fF^i_{yf}-l^i_rF^i_{yr}}{I^i_z}
        \end{equation}
    \end{subequations}
where $\xi^i = \begin{bmatrix}\xi_x^i\\\xi_y^i\end{bmatrix}$ and $\xi_x^i$, $\xi_y^i$ are the global longitudinal and lateral positions of $i^{th}$ AV coordinate respectively, $m^i$ is the mass of AV, $l^i_f$ and $l^i_r$ are the distances to the front and rear axles from the center of gravity respectively, $\psi^i$ is the yaw angle, $\omega^i$ is the yaw rate, $F^i_{yf}$ and $F^i_{yr}$ are the lateral forces of the front and the rear wheels respectively, $I^i_z$ is the moment of inertia, $a^i = \Big(\frac{F^i_{xf}+F^i_{xr}}{m^i}\Big)$ is the longitudinal acceleration, and $v^i_x$ and $v^i_y$ are the longitudinal and lateral velocities of the $i^{th}$ AV respectively.
    
Considering the vehicle heading angle (yaw angle) is small during the lane changing scenario implies that $\cos{\psi^i}\approx 1$ and $\sin{\psi^i}\approx \psi^i$. Therefore \eqref{eq:dyn_long_lat_position} can be rewritten as
        \begin{equation}
            \label{eq:dyn_long_lat_position_simpli}
            \Dot{\xi}^i = \begin{bmatrix}\Dot{\xi}_x^i\\\Dot{\xi}_y^i \end{bmatrix} = \begin{bmatrix}1&-\psi^i\\\psi^i&1 \end{bmatrix}\begin{bmatrix}v_x^i\\v_y^i\end{bmatrix}
        \end{equation}
The lateral tire force of the front and rear wheels of AV can be written as
    \begin{subequations}
        \label{eq:dyn_tire_force}
        \begin{equation}
            \label{eq:dyn_tire_force_front}
            F^i_{yf} = 2C^i_f\Big(\delta^i-\dfrac{v^i_y+l^i_f\omega^i}{v^i_x}\Big),
        \end{equation}
        \begin{equation}
            \label{eq:dyn_tire_force_rear}
            F^i_{yr} = -2C^i_r\dfrac{v^i_y-l^i_r\omega}{v^i_x}.
        \end{equation}
    \end{subequations}
where $C^i_f$ and $C^i_r$ are the cornering stiffness of the front and rear axle, respectively, and $\delta^i$ is the steering angle of the front wheel. Substitute \eqref{eq:dyn_long_lat_position_simpli} and \eqref{eq:dyn_tire_force} into \eqref{eq:dynamic_model_NL}, we can now reformulate the vehicle dynamic model into a state-space representation as follows:
    \begin{equation}
        \label{eq:ss_cont_dynamics}
        \Dot{x}^i(t) = Ax^i(t) + Bu^i(t).
    \end{equation}
where the state vector, $x^i(t) = [\xi^{i^\top},~v^i_x,~v^i_y,~\psi^i,~\omega^i]^\top\in\mathbb{R}^6$, the control input vector, $u^i(t) = [a^i,~\delta^i]^\top\in\mathbb{R}^2$, the system matrix,
    
$A = \begin{bmatrix}
        0&0&1&0&-v^i_y&0\\
        0&0&0&1&v^i_x&0\\
        0&0&0&0&0&v^i_y\\
        0&0&0&-2\frac{C^i_f+C^i_r}{m^iv^i_x}&0&2\frac{C^i_rl^i_r-C^i_fl^i_f}{m^iv^i_x}-v^i_x\\
        0&0&0&0&0&1\\
        0&0&0&2\frac{C^i_rl^i_r-C^i_fl^i_f}{I^i_zv^i_x}&0&-2\frac{C^i_rl^{i^2}_r+C^i_fl^{i^2}_f}{I^i_zv^i_x}
    \end{bmatrix}$,
and the input matrix, $B = \begin{bmatrix}
        0&0&1&0&0&0\\
        0&0&0&\frac{2C^i_f}{m^i}&0&\frac{2C^i_fl^i_f}{I^i_z}
    \end{bmatrix}^\top$.\\

We discretize \eqref{eq:ss_cont_dynamics} using the \emph{zero-order hold} method with sampling time $t_s$ to get the discrete state space representation of system dynamics \eqref{eq:ss_cont_dynamics} for further analysis,
    \begin{equation}
        \label{eq:ss_dist_dynamics}
        x^i_{k+1} = A_dx^i_k + B_du^i_{k}.
    \end{equation}
where the discrete state vector, $x^i_k = [\xi^{i^\top}_k,~v^i_{x_k},~v^i_{y_k},~\psi^i_k,~\omega^i_k]^\top$, the discrete control input vector, $u^i_k = [a^i_{k},~\delta^i_k]^\top$, and $A_d$ and $B_d$ are the discretized system matrix and input matrix respectively. The discrete states $x^i_k$ and input $u^i_k$ of each AV are considered to be bounded by their upper and lower limits as follows:
    \begin{multline}
        \label{eq:state_input_constraints}
        \begin{bmatrix}\underline{p}_x\\\underline{p}_y \end{bmatrix}\leq \xi^i_k \leq \begin{bmatrix}\overline{p}_x\\\overline{p}_y \end{bmatrix},~
        \underline{v}_x\leq v_{x_k}^i\leq \overline{v}_x,~ 
        \underline{v}_y\leq v_{y_k}^i\leq \overline{v}_y,\\
        \underline{\psi}\leq \psi^i_k\leq\overline{\psi},~
        \underline{\omega}\leq \omega^i_k\leq\overline{\omega},~
        \underline{a}\leq a^i_{k}\leq\overline{a},~
        \underline{\delta}\leq \delta^i_k\leq\overline{\delta}.
    \end{multline}

Further, a safe distance should be maintained to prevent collision between vehicle $i$ and the neighboring vehicle $j$. Such a distance can be modeled as:
    \begin{equation}
        \label{eq:safe_distance_const}
        \Vert \xi^i_{k} - \xi^j_{k}\Vert \geq \gamma.
    \end{equation}
where $\gamma$ is user defined minimum safe distance; however, the collision avoidance constraint \eqref{eq:safe_distance_const} is nonlinear and non convex in nature. Therefore, a linear approximation of \eqref{eq:safe_distance_const} is required to design a convex control invariant terminal set. The linear approximation of \eqref{eq:safe_distance_const} and the design of the terminal set will be discussed in a later section. 

We posit that vehicles may cross the junction during the active green interval of the traffic signal upon their entry into the \emph{control zone}. However, it is essential to understand that this does not imply that all vehicles must traverse the intersection during the same green phase upon reaching the junction point. They might pass the junction in the upcoming green interval, which causes congestion at the junction point due to heavy traffic demands. Therefore, we are interested in addressing the following problem.
    \begin{prob}
        Consider a signalized junction ($SJ$). AVs are entering the zone at random times. Design a feedback control policy for each AV such that the accumulation of vehicles during the red signal phase is minimized to avoid congestion. Also, they can perform safe lane-changing maneuvers inside the \emph{control zone} if required. 
    \end{prob}

\section{MPC Formulation for Control Action for AVs} 
\label{sec:req_comp_mpc}
Each AV is required to attain a certain velocity to reach and stop at the junction at a given time during the red signal interval. Also, it is imperative to determine appropriate constraints to formulate the constrained convex optimization problem. Furthermore, developing a terminal set that ensures the feasibility and convergence properties of MPC is crucial.

\subsection{Reference Velocity Generation}
\label{subsec:ref_vel_gen}
The proposed algorithm in Algorithm \hyperref[algo:algorithm_1]{1}, for generating reference velocity, is similar to \cite{R_Sen_2024}. However, unlike \cite{R_Sen_2024}, here, the reference velocity for AV is generated to reach and stop at the junction point at a given time during the red signal interval. The algorithm proposed here generates a reference velocity for sequential coordination of AVs to stop at the junction point during a red signal interval such that \emph{congestion} or \emph{oversaturation} at the junction point can be avoided.

Assume that every signalized junction ($SJ$) has a control zone of its own. An AV can obtain data regarding the duration of the traffic signal of that $SJ$ after it has entered this control zone. Consider that $d_m$ is the position of AV from the upcoming $SJ_m$. Assume that $SJ_m$ transmits the following information to an AV upon its entry into the zone: \textbf{(i)} the time interval $\tau_0$, after which the subsequent signal, either green or red, will activate, \textbf{(ii)} the durations of red time $\tau_r$ and green time $\tau_g$, and \textbf{(iii)} the unique ID $nV$, assigned to that AV upon its entry. We define \emph{critical density} as the density of the traffic at which the traffic flow shifts to a congested condition from a free-flow condition. However, it is worth noting that \emph{critical density} is a user-defined parameter.

    \begin{algorithm}
        \small{\caption{Generation of Reference velocity ($v_{x_{r}}$)}
        \textbf{Input:} $d_m$, $nV$, $[\underline{v}_x,\hspace{2pt} \overline{v}_x]$, $\tau_d$, $\tau_{0}$, $\tau_g$, $\tau_r$ \\
        \textbf{Output:} Reference longitudinal velocity, $v_{x_{r}}$
        \begin{algorithmic}[1]
            \If{green signal ON after $\tau_0$}
                \State $\tau_{r_{1}} \gets (\tau_{0} + \tau_g)$, $\tau_{g_{2}} \gets (\tau_{r_{1}} + \tau_r)$, \ldots \text{up-to} $\Gamma$
                \State $\mathcal{N} \gets \{\tau_0,\tau_{r_{1}},\tau_{g_{2}}, \ldots \text{up-to}\hspace{3pt} \Gamma\}$
            \ElsIf{red signal ON after $\tau_0$}
                \State $\tau_{g_{2}} \gets (\tau_0 + \tau_r)$, $\tau_{r_{2}} \gets (\tau_{g_2} + \tau_g)$\ldots \text{up-to} $\Gamma$
                \State $\mathcal{N} \gets \{\tau_0,\tau_{g_{2}},\tau_{r_{2}}, \ldots \text{up-to}\hspace{3pt} \Gamma\}$
            \EndIf
            \For {$\mathfrak{j}$ = $1$ : \texttt{length}$(\mathcal{N})$}
                \State $V_{feasible} \gets \Big[\dfrac{d_m}{\tau_{r_{\mathfrak{j}}}-\tau_d},\dfrac{d_m}{\tau_{g_{\mathfrak{j}}}+\tau_d}\Big]\cap [\underline{v_x},\hspace{2pt}\overline{v}_x]$
                \vspace{1pt}
                \If{($V_{feasible} \neq \emptyset$)}
                    \State $[v_{x_{lt}},v_{x_{rt}}] \gets V_{feasible}$
                    \State \texttt{break}
                \Else
                    \If{($nV\leq$ \emph{critical density})}
                        \State $t_{reach} \gets \dfrac{1}{2}(\tau_{r_\mathfrak{j}} + \tau_{g_{\mathfrak{j}+1}})$
                        \State $v_{x_{rt}} \gets d_m/t_{reach}$
                    \Else
                        \State $\mathfrak{j} \gets \mathfrak{j}+1$
                        \State $t_{reach} \gets \dfrac{1}{2}(\tau_{r_\mathfrak{j}} + \tau_{g_{\mathfrak{j}+1}})$
                        \State $v_{x_{rt}} \gets d_m/t_{reach}$
                    \EndIf
                    \State \texttt{break}
                \EndIf
            \EndFor\\
            \Return $v_{x_{r}} \gets v_{x_{rt}}$
        \end{algorithmic}}
        \label{algo:algorithm_1}
    \end{algorithm}
    \textbf{Explanation for Algorithm \hyperref[algo:algorithm_1]{1}:}
    \emph{Steps 1-7:} Based on the data ($\tau_0$, $\tau_g$, $\tau_r$, and the activation of either the red or green light following $\tau_0$ seconds) obtained from the $m^{th}$ junction $SJ_m$, a series of time periods is obtained. In \emph{steps} $1$ through $3$, we calculate $\tau_{r_1} = \tau_0 + \tau_g$, $\tau_{g_2} = \tau_{r_1} + \tau_r$ up-to $\Gamma$, to construct a sequence $\mathcal{N}$ if signal turns to green after $\tau_0$s. Similarly, in \emph{steps} $4$ through $6$, we calculate $\tau_{g_2} = \tau_0 + \tau_r$, $\tau_{r_2} = \tau_{g_2} + \tau_g$ up-to $\Gamma$, to construct a sequence $\mathcal{N}$ if signal turns to red after $\tau_0$s. As we have considered a heavy traffic scenario, unlike \cite{R_Sen_2024}, the time horizon $\Gamma$ does not need to be considered large enough (possibly: $\Gamma \geq (\tau_0 + \tau_r + \tau_g) $). For instance, let $SJ_m$ broadcasts: $\tau_0 = 10$ s, $\tau_g = 10$ s, and $\tau_r = 30$ s, and the red signal turns ON after $\tau_0$ sec. Then, by selecting $\Gamma = 90$ s, the generated signal sequence is: $\mathcal{N} = \{0,~10,~40,~50,~80,~90\}$.
    
    \vspace{2pt}
    
    \emph{Steps 8-23:} In the following steps, we compute the longitudinal velocity for an AV. In \emph{step} $9$ we compute the feasible longitudinal velocity interval of an AV considering the constraint on $v_{x_k}$ as follows: $V_{feasible} = \Big[\dfrac{d_m}{\tau_{r_{j}}-\tau_d},\dfrac{d_m}{\tau_{g_{j}}+\tau_d}\Big]\cap [\underline{v}_x,\hspace{2pt}\overline{v}_x]$, where $\tau_d$ is a positive number used as a safety margin. If feasible set $V_{feasible}$ is not empty, then the highest value of the set, $v_{x_{rt}}$, is considered the reference longitudinal velocity. However, if $V_{feasible} = \emptyset$, due to heavy traffic conditions, AV has to stop at the upcoming red signal. To avoid congestion at the junction point and improve the traffic flow, sequential coordination among AVs is needed. In dense traffic, the surges of traffic propagate backward to maintain a safe distance by frequent accelerating and decelerating. It is therefore required that the number of vehicles at the junction point should be less than or equal to \emph{critical density} to reduce the unnecessary wait time for the commuter near $SJ$. As each AV is assigned with an index number ($nV$) upon entering the zone, if $nV \leq \emph{critical density}$, then they compute the velocity $v_{x_r}$ to reach and stop at the junction point at the upcoming red signal, as shown in \emph{steps} $15-16$. If $nV > \emph{critical density}$, then AVs adjust their velocity to reach the $SJ$ at the next cycle of the red phase and increase their headway distance before facing the jam, as shown in \emph{steps} $18-20$. This calculated $v_{x_{rt}}$ is considered as the reference longitudinal velocity $v_{x_{r}}$.

Upon the generation of $v_{x_{r}}$ for an AV utilizing Algorithm \hyperref[algo:algorithm_1]{1}, it is necessary to implement a suitable control system to attain the specified velocity profile such that the vehicle can reach and stop at the junction point. We use the MPC framework to address the \emph{oversaturation} scenario and the lane-changing maneuver in heavy traffic. We considered that each AV has its own sensing equipment and local controllers to generate appropriate control signals. 

\subsection{Linear Approximation of Collision Avoidance Constraint}
\label{subsec:linear_approx_const}
To avoid potential accidents between the AVs on the road, each of them is required to maintain a minimum safe distance from the other, which we modeled as a Euclidean distance between AVs that must be greater than or equal to the minimum safe gap, shown in \eqref{eq:safe_distance_const}, which is nonlinear. In order to implement the linear MPC and to ensure the convexity of the invariant set, it is imperative that all constraints are linear and convex in nature. In the following Algorithm \hyperref[algo:algorithm_linearapprox]{2}, we draw up the linear approximation of the collision avoidance constraint \eqref{eq:safe_distance_const}. 

\begin{figure*}[thbp]
      \centering
      \includegraphics[height = 5.0cm, width = 18.1cm]{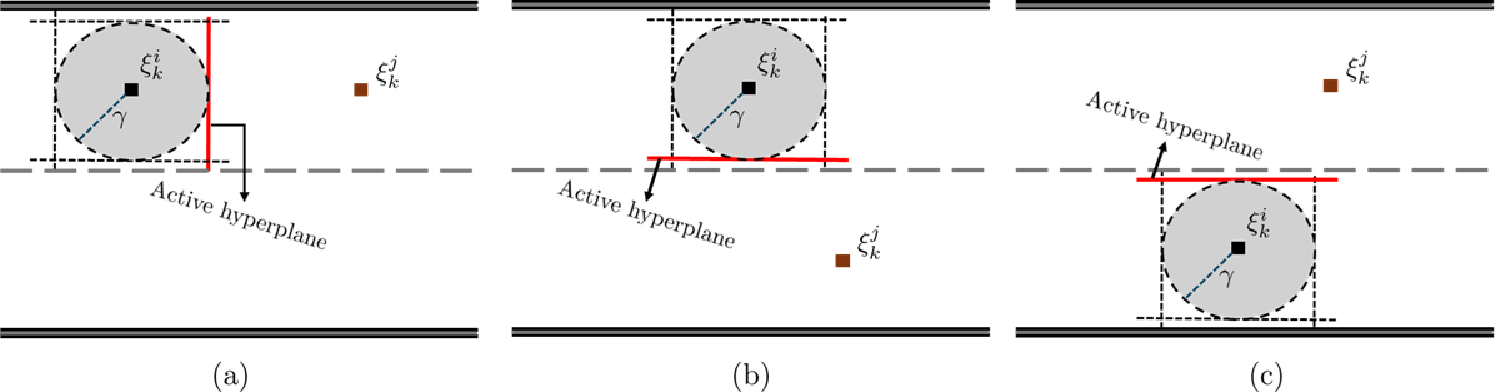}
      \caption{Linearized collision avoidance constraint: (a) The active hyperplane (solid red) when $j^{th}$ AV is ahead on the same lane, (b) The active hyperplane when $j^{th}$ AV is on the right lane, (c) The active hyperplane when $j^{th}$ AV is on the left lane.}
      \label{fig:linearized_collision_avoidance}
   \end{figure*}

The Euclidean distance depicted in \eqref{eq:safe_distance_const} essentially represents a circle centered at $i^{th}$ AV. We outer approximate this circle by a square, whose sides are generally hyperplanes: $w^\top_l(\xi^j_k-\xi^i_k) = \gamma$, and $i^{th}$ AV lies inside the area enclosed by halfspaces: $w^\top_l(\xi^j_k - \xi^i_k) \leq \gamma$.
Here $w_l = [\cos{\frac{2\pi l}{4}},~\sin{\frac{2\pi l}{4}}]^\top$ with $l = 0,\ldots,3$, $\gamma$ is the user-defined minimum safe distance that is the radius of the circle centered at $i^{th}$ AV. Unlike \cite{M_Farina_2015} here, we use $4$ hyperplanes to out-approximate the circle as shown in Fig.~(\ref{fig:linearized_collision_avoidance}), thereby reducing computational time. The more hyperplanes we use, the more accurate the circle approximation will be, but the computational time increases.

    \begin{algorithm}[H]
        \caption{Linear approximation of Collision Avoidance}
        \begin{algorithmic}[1]
            \Require $\gamma$, $\{\xi^i_{p|k}\}_{p=0}^N$, $\{\xi^j_{p|k}\}_{p=0}^N$, \text{sensor range} ($R$), $N$.
            \Ensure normal vector, $w_l$
            \For{$l = 0$ : $3$}
              \State Define normal vector: $w_l \gets \left[\begin{smallmatrix}\cos(2\pi l/4)\\\sin(2\pi l/4)\end{smallmatrix}\right]$
            \EndFor
            \If{$\Vert \xi^j_k - \xi^i_k\Vert \leq R$}
                $j\rightarrow$\text{neighbouring AV of $i^{th}$ AV}
                \For{\text{each} $j$}
                    \For{$p = 0 :N$}
                        \State $p_x \gets (\xi^j_{X_{p|k}}- \xi^i_{X_{p|k}})$,~~~ $p_y \gets (\xi^j_{Y_{p|k}}- \xi^i_{Y_{p|k}})$
                        \State $d_\epsilon\rightarrow$ small positive constant.
                        \If{$p_x > 0$ and $\vert p_y\vert \leq d_\epsilon$}
                            \State $w_l \gets w_0$\hspace{3.5em}\text{\% $j^{th}$ AV on same lane}
                        \ElsIf{($p_x > 0$ or $p_x = 0$) and $p_y > d_\epsilon$}
                            \State $w_l \gets w_1$\hspace{4em}\text{\% $j^{th}$ AV on left lane}
                        \ElsIf{($p_x > 0$ or $p_x = 0$) and $p_y < -d_\epsilon$}
                            \State $w_l \gets w_3$\hspace{4em}\text{\% $j^{th}$ AV on right lane}
                        \ElsIf{$p_x < 0$} \hspace{3.0em}\text{\%$j^{th}$ AV is behind}
                            \State \text{No requirement for active hyperplanes}
                        \EndIf
                    \EndFor
                \EndFor
            \EndIf
        \end{algorithmic}
        \label{algo:algorithm_linearapprox}
    \end{algorithm}

The linear approximation of the collision avoidance constraint \eqref{eq:safe_distance_const} can be written as:
    \begin{equation}
        \label{eq:safe_distance_const_linear}
        w^\top_{l}(\xi^j_k-\xi^i_k) \geq \gamma.
    \end{equation}
Note that each vehicle exchanges its future predicted information only with neighboring vehicles via V2V communication. We also consider that each AV is equipped with a proximity sensor (like LiDAR). If a vehicle enters within the range of an AV's proximity sensors, it is considered to be a neighboring vehicle.

We can, therefore, rewrite the state and input constraint mentioned in \eqref{eq:state_input_constraints} and \eqref{eq:safe_distance_const_linear} in the following form:
    \begin{equation}
        \label{eq:all_constraint_together}
        H^i_{x_k} x^i_k \leq g^i_{x_k},~~~ H^i_u u^i_k \leq g^i_u.
    \end{equation}
where 
$H^i_{x_k} = \begin{bmatrix}C_{x_k}\\F_{x_k}\end{bmatrix}$. $\mathcal{V}^i$ denotes the set of neighbouring vehicles of $i^{th}$ AV, 
$C_{x_k} = \left[\begin{smallmatrix}c^1_{x_k}\\\vdots\\c^j_{x_k}\end{smallmatrix}\right]$ for $j\in\mathcal{V}^i$ and 

$c^m_{x_k} = [\cos{\frac{2\pi l}{n}}~~ \sin{\frac{2\pi l}{n}}~~ 0~~ 0~~ 0~~ 0]$ where $m\in j$ and 

$F_{x_k} = \left[\begin{smallmatrix}
    1 & 0 & 0 & 0 & 0 & 0\\
    0 & 1 & 0 & 0 & 0 & 0\\
    0 & 0 & 1 & 0 & 0 & 0\\
    0 & 0 & 0 & 1 & 0 & 0\\
    0 & 0 & 0 & 0 & 1 & 0\\
    0 & 0 & 0 & 0 & 0 & 1\\
    -1 & 0 & 0 & 0 & 0 & 0\\
    0 & -1 & 0 & 0 & 0 & 0\\
    0 & 0 & -1 & 0 & 0 & 0\\
    0 & 0 & 0 & -1 & 0 & 0\\
    0 & 0 & 0 & 0 & -1 & 0\\
    0 & 0 & 0 & 0 & 0 & -1
\end{smallmatrix}\right]$, $H^i_u = \begin{bmatrix}1 & 0\\0 & 1\\ -1 & 0\\ 0 & -1\end{bmatrix}$,

$\mspace{-15mu}g^i_{x_k} = [(-\gamma + w^\top_{l}\xi^j_k), \overline{p}_x, \overline{p}_y, \overline{v}_x, \overline{v}_y, \overline{\psi}, \overline{\omega}, \underline{p}_x, \underline{p}_y, \underline{v}_x, \underline{v}_y, \underline{\psi}, \underline{\omega}]^\top$,

$g^i_u = [\overline{a}~~~\overline{\delta}~~~\underline{a}~~~\underline{\delta}]^\top$. 

The constraints mentioned in \eqref{eq:all_constraint_together} are the halfspace representation of state constraint set $\mathscr{X}^i$ and input constraint set $\mathscr{U}^i$, respectively.

\subsection{Computation of Terminal Set} 
\label{subsec:CI_set}
In the dual-mode MPC strategy \cite{F_Borrelli_2017}, the concept of the terminal set suggests that at each time instant after the end of the prediction horizon $N$, the predicted system state $x^i_{N|k}$ should reach the terminal set $\mathscr{X}^i_{\mathbb{T}_k}\subseteq \mathscr{X}^i$. The set $\mathscr{X}^i_{\mathbb{T}_k}$ is considered to be a forward-time invariant set. In this article, we have designed a \emph{maximal control invariant set} $\mathscr{X}^i_{\mathbb{T}_k}$. According to the definition \cite[Def. 10.9 and 10.10]{F_Borrelli_2017}, a maximal control invariant set is a control invariant set where exists a closed loop control $u^i_k \in \mathscr{U}^i$ inside $\mathscr{X}^i_{\mathbb{T}_k}$ such that system state $x^i_{k+1} \in \mathscr{E}^i_{\mathbb{T}_k}$ if $x^i_k \in \mathscr{X}^i_{\mathbb{T}_k}$ and it is a union of all the control invariant sets existed in $\mathscr{X}^i$. In contrast to the terminal set designed in \cite{R_Sen_2024}, a control invariant terminal set does not pertain to a specific control structure; instead, it is formulated for any closed-loop control that adheres to the control constraint condition and maintains the system state within the set for future time. Consequently, the \emph{maximal control invariant set} is regarded as a superset of \emph{the maximal positive invariant set}. 

It follows from the theorem in \cite[Theorem 10.1]{F_Borrelli_2017} that $\mathscr{X}^i_{\mathbb{T}_k}$ is said to be maximal control invariant if $\mathscr{X}^i_{\mathbb{T}_k} \subseteq \text{Pre}(\mathscr{X}^i_{\mathbb{T}_k})$ where $\text{Pre}(\mathscr{X}^i_{\mathbb{T}_k})$ is a precusor set of $\mathscr{X}^i_{\mathbb{T}_k}$\cite[Def. 10.2]{F_Borrelli_2017}. This indicates that $\text{Pre}(\mathscr{X}^i_{\mathbb{T}_k})\cap\mathscr{X}^i_{\mathbb{T}_k} = \mathscr{X}^i_{\mathbb{T}_k}$. The following Algorithm \hyperref[algo:algorithm_terminal]{3} delineates the computation of $\mathscr{X}^i_{\mathbb{T}_k}$. To compute $\mathscr{X}^i_{\mathbb{T}_k}$, we initialize a set $\mathcal{S}_0 = \mathscr{X}^i$. The precursor set of $\mathcal{S}_0$ is defined as:
    \begin{equation}
        \label{eq:precursor_set}
        \begin{split}
            \text{Pre}(\mathcal{S}_0) &= \{x^i_k : \exists u^i_k\in\mathscr{U}^i ~\text{such that}~ x^i_{k+1}\in\mathcal{S}_0\}\\
        &= \{x^i_k : A_dx^i_k = x^i_{k+1} + (-B_du^i_k), ~\forall u^i_k\in\mathscr{U}^i,\\ &~~~~~~~~~~~~~~~~~~~~~~~~~~~~~~~~~~~~~~~~~~x^i_{k+1}\in\mathcal{S}_0\}\\
        &= \{x^i_k : A_dx^i_k \in \mathcal{S}_0 \oplus ((-B_d)\circ\mathscr{U}^i)\}\\
        &= \{x^i_k : x^i_k \in (\mathcal{S}_0 \oplus ((-B_d)\circ\mathscr{U}^i))\circ A_d\}.
        \end{split}
    \end{equation}
Similarly, in the subsequent algorithm, the precursor set of $\mathcal{S}_h$ at each iteration is calculated as described in \eqref{eq:precursor_set} until the set converges to $\mathscr{X}^i_{\mathbb{T}_k}$. 
    \begin{algorithm}
        \small{\caption{Computation of $\mathscr{X}^i_{\mathbb{T}_k}$ \cite[Chapter 10]{F_Borrelli_2017} }
        \textbf{Input:} (i) System dynamics \eqref{eq:ss_dist_dynamics}, (ii) State and input constraint set, $\mathscr{X}^i$ and $\mathscr{U}^i$ as in \eqref{eq:all_constraint_together}, respectively. \\
        \textbf{Output:} Maximal control invariant set $\mathscr{X}_{\mathbb{T}_k}^i$
        \begin{algorithmic}[1]
            \State $\mathcal{S}_0 \gets \mathscr{X}^i$, $h \gets 0$
            \While {(1)}
                \State $\mathcal{S}_{h+1} \gets \text{Pre}(\mathcal{S}_h)\cap \mathcal{S}_h$
                \If{($\mathcal{S}_{h+1} = \mathcal{S}_h$)}\hspace{3pt}\texttt{break}
                \Else
                    \State $\mathcal{S}_h \gets \mathcal{S}_{h+1}$
                \EndIf
                    \State $h \gets h+1$
            \EndWhile\\
            \Return $\mathscr{X}_{\mathbb{T}_k}^i \gets \mathcal{S}_h$
        \end{algorithmic}
        \label{algo:algorithm_terminal}}
    \end{algorithm}
  
\subsection{Cost Function and MPC formulation} 
\label{sec:cost_func}
In this article, we are solving a set-point tracking problem. For this we have considered a reference point $x^i_{r} = [\xi^i_{x_r},~\xi^i_{y_r},~v^i_{x_{r}},~0,~0,~0]^\top$. We define the quadratic reference tracking stage cost function as follows: $\Vert x^i_{p|k} - x^i_{r}\Vert^2_Q + \Vert u^i_{p|k} \Vert^2_R$ for $p = 0,\ldots, N-1$, which regulates the cost until prediction horizon $N$. Here, $  Q\succeq 0$ and $R\succ 0$ represent the stage cost weight matrices for state and control input, respectively. Additionally, we incorporate a terminal cost $\Vert x^i_{N|k} -x^i_{r} \Vert^2_P$ to ensure the feasibility and stability of MPC, where $P \succ 0$ denotes the terminal cost weight matrix.

Therefore, we can now formulate the constrained optimization problem $\mathbb{OP}$ as follows:
    \begin{subequations}
        \label{eq:MPC_framework}
        \begin{multline}
            \label{eq:MPC_cost}
            \mathbb{OP}:\\\min_{\mathcal{U}_{k}^i \in \mathscr{U}^i} J_{k}^i \triangleq \Vert x^i_{N|k} -x^i_{r} \Vert^2_P+ \sum_{p = 0}^{N-1}\Vert x^i_{p|k} - x^i_{r}\Vert^2_Q + \Vert u^i_{p|k} \Vert^2_R
        \end{multline}
        \begin{equation}
            \label{eq:system_dynamics}
            \text{subject to,} ~~~~x_{p+1|k}^i = A_dx_{p|k}^i + B_du_{p|k}^i, ~~~ \forall p \in \mathbb{I}_0^{N-1}
        \end{equation}
        \begin{equation}
            \label{eq:state_control_constraint_MPC}
            ~~~~~~~~~~~~~~~~~H^i_{x_k}x^i_{p|k} \leq g_{x_k}^i,~~ H^i_uu^i_{p|k} \leq g^i_u, ~~ \forall p \in \mathbb{I}_0^{N-1}
        \end{equation}
        \begin{equation}
            \label{eq:initial_condition_MPC}
            x_{0|k}^i = x_{k}^i,
        \end{equation}
        \begin{equation}
            \label{eq:terminal_constraint_MPC}
            x_{N|k}^i\in\mathscr{X}_{\mathbb{T}_k}^i.
        \end{equation}
    \end{subequations}
The RHC approach in the MPC framework is employed to generate a sequence of control inputs for the $i^{th}$ AV. This approach necessitates the optimization $\mathbb{OP}$ to be solved repeatedly at each time step over the prediction horizon $N$. It is conceivable that the problem $\mathbb{OP}$ becomes infeasible at a certain time step; hence, a terminal constraint \eqref{eq:terminal_constraint_MPC} is integrated to ensure the feasibility of $\mathbb{OP}$.

\subsection{Lane Change Maneuver} 
\label{subsec:lane_change}
\begin{figure}[thbp]
      \centering
      \includegraphics[height = 2.2cm, width = 8.7cm]{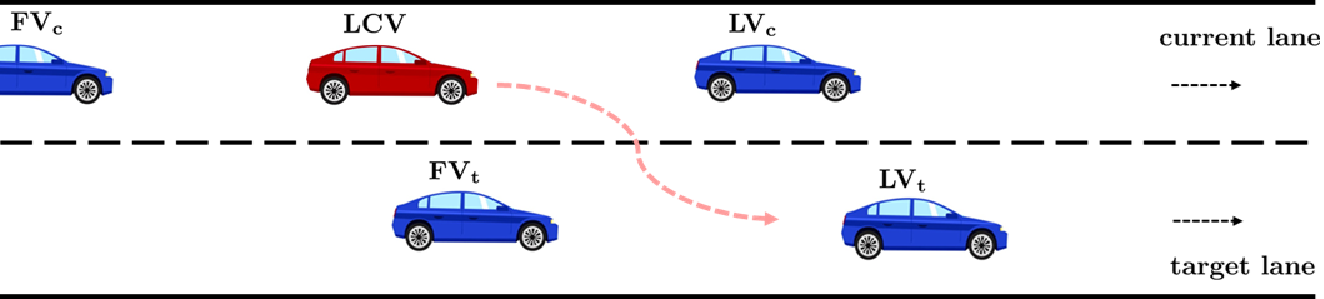}
      \caption{Lane change scenario}
      \label{fig:lane_change_setting}
   \end{figure}

Accidents often occur during lane changes because this complex movement demands acute awareness of the surroundings and the capacity to judge distances and speeds in order to prevent collisions. Consequently, it is essential in the context of automated driving. This subsection will address the cooperative lane change maneuver in an autonomous traffic environment.  Fig.~(\ref{fig:lane_change_setting}) illustrates lane change configuration in an autonomous traffic environment, where lane changing vehicle (\textbf{LCV}) represents the vehicle intending to change lanes, vehicles $\mathbf{FV_c}$ and $\mathbf{FV_t}$ are the following vehicles in the current and target lanes respectively, and $\mathbf{LV_c}$ and $\mathbf{LV_t}$ signify the leading vehicles in the current and target lanes respectively.

We are examining a high-traffic situation in which an autonomous vehicle intends to shift lanes. Since only a single lane change movement is included, for simplicity, Fig.~(\ref{fig:lane_change_setting}) depicts two lanes instead of three. Furthermore, we are presuming that only a single vehicle is capable of executing the lane-changing action at any given moment. A similar principle can be applied to a scenario in which more than one vehicle is changing lanes. The following Algorithm \hyperref[algo:algorithm_lanechange]{4} demonstrates the cooperative lane change configuration.

    \begin{algorithm}[H]
        \caption{Cooperative Lane Change Maneuver of \textbf{LCV}}
        \begin{algorithmic}[1]
        \State Compute inter-vehicle distance:
        $D \gets \| \xi^{lv_t}_k - \xi^{fv_t}_k \|$
        \If{$D > 2\gamma$ \textbf{(i.e., no AV detected within sensor radius at the target lane)}}
            \State \textbf{LCV} initiate lane change by updating the $y$-coordinate 
            \Statex\hspace{1.5em}to new $\xi^{lcv}_{y_{rn}}$.
            \If{(no AV in the sensor's range of \textbf{LCV} at the target lane)}
                \State The \textbf{LCV} treats the junction stop line as a hard 
                \Statex\hspace{3em} constraint if the signal is red.
            \ElsIf{(During lane change, if $\mathbf{FV_t}$ or $\mathbf{LV_t}$ or both enters the sensor range)}
                \State \textbf{LCV} establishs communication with both vehicles.
                \State Receives predicted positional data:
                    \[
                    \{\xi^{fv_t}_{p|k}\}_{p=0}^{N},~~
                    \{\xi^{lv_t}_{p|k}\}_{p=0}^{N}
                    \]
                \State Transmits to $\mathbf{FV_t}$ the predicted and anticipated 
                \Statex\hspace{3em}positions of \textbf{LCV}:
                    \[
                    \{\xi^{lcv}_{p|k}\}_{p=0}^{N}, \quad 
                    \{\overline{\xi}^{lcv}_{p|k}\}_{p=0}^{N}
                    \]
                \State Computes the anticipated positions as:
                    \[
                    \overline{\xi}^{lcv}_{p|k} = \xi^{lv_t}_{p|k} - 
                    \begin{bmatrix}
                    \gamma \\ 0
                    \end{bmatrix}
                    =
                    \begin{bmatrix}
                    \xi^{lv_t}_{x_{p|k}} - \gamma \\
                    \xi^{lcv}_{y_{rn}}
                    \end{bmatrix}
                    \]
                \State \textbf{Repeat} Steps 7–10 until:
                    \[
                    |\xi^{lcv}_{y_{p|k}} - \xi^{lcv}_{y_{rn}}| \leq \epsilon_{tol}
                    \]
                \Statex \hspace{2em} where $\epsilon_{tol}$ indicates the tolerance value. 
            \EndIf
        \Else 
            \State \textbf{LCV} continues on the same lane.
        \EndIf   
        \end{algorithmic}
        \label{algo:algorithm_lanechange}
    \end{algorithm}

    \textbf{Explanation for Algorithm \hyperref[algo:algorithm_lanechange]{4}:} \emph{Steps 1-3:}
    The \textbf{LCV} initially checks for neighbouring AVs in the target lane. If no AVs are detected, that is, if there is enough space for a lane change, \textbf{LCV} commences the lane change towards the designated lateral position. The new $y$-coordinate is the center line $y$-coordinate of the target lane, which is also a new reference point $\xi^{lcv}_{y_{rn}}$ of $\mathbf{LCV}$ (step $3$).
    
     \emph{Steps 6-11:} During the maneuver, once $\mathbf{FV_t}$ and $\mathbf{LV_t}$ enters the \textbf{LCV}'s sensor region, it establishes communication with them. In step $8$, at each time instant $\mathbf{LCV}$ receives predicted positional data from $\mathbf{FV_t}$ and $\mathbf{LV_t}$, which are generated by executing the $\mathbb{OP}$. In steps $9-10$, $\mathbf{LCV}$ transmits its current predicted positional data, $\{\xi^{lcv}_{p|k}\}_{p=0}^{N}$ obtained by executing the $\mathbb{OP}$ to $\mathbf{FV_t}$ only. Additionally, $\mathbf{LCV}$ transmits its anticipated final positional data, $\{\overline{\xi}^{lcv}_{p|k}\}_{p=0}^{N}$ to $\mathbf{FV_t}$. Both the current and anticipated positional data of $\mathbf{LCV}$ act as collision avoidance constraints to $\mathbf{FV_t}$, which helps to generate a suitable terminal set. In step $11$, steps $7$ to $10$ required to repeat until $\vert \xi^{lcv}_{y_{p|k}} - \xi^{lcv}_{y_{rn}}\vert \leq \epsilon_{tol}$ achieved. This condition indicates that the \textbf{LCV} has integrated into the target lane between $\mathbf{LV_t}$ and $\mathbf{FV_t}$.
\begin{rem}
    As $\mathbf{FV_t}$ receives anticipated final position data sequence $\{\overline{\xi}^{lcv}_{p|k}\}_{p=0}^{N}$ at each time step from \textbf{LCV}, it act as a new collision avoidance constraint \eqref{eq:safe_distance_const_linear} on $\mathbf{FV_t}$. This enables $\mathbf{FV_t}$ to compute a more restricted terminal set, thereby reducing its velocity and ensuring a sufficient gap between $\mathbf{LV_t}$ and $\mathbf{FV_t}$. The generation of a terminal set guarantees the seamless functioning of each autonomous vehicle at every time step, and the exchange of positioning data with neighboring vehicles is adequate for safe operation.
\end{rem}

\section{Recursive Feasibility and Stability Analysis}
\label{sec:feasible_convergence}
This section addresses the theoretical analysis of a closed-loop system under the MPC feedback technique. In the subsequent lemmas, we demonstrate that the imposition of constraint \eqref{eq:terminal_constraint_MPC} will guarantee the feasibility and stability of $\mathbb{OP}$ at each stage.
    
\subsection{Recursive Feasibility}
\label{subsec:feasibility}
\begin{lem}
    \label{lemma:lemma_1}
    Assume that there exists a nonempty terminal set $\mathscr{X}^i_{\mathbb{T}_k}\subseteq\mathscr{X}^i$ such that the predicted terminal state $x^i_{N|k}\in\mathscr{X}^i_{\mathbb{T}_k}$. If the optimization $\mathbb{OP}$ is feasible at time step $k$, then it will be feasible for any future time step $k+p$, $\forall p\in\mathbb{I}^\infty_1$.
\end{lem}
\begin{IEEEproof}
    In order to show the feasibility of $\mathbb{OP}$, it is essential to establish that, for any given $(x^i_k, u^i_k )\in \mathscr{X}^i\times\mathscr{U}^i$ at time step $k$, there exist a solution at time step $k+1$. Since $\mathbb{OP}$ is feasible at time $k$, there exists an optimal solution of \eqref{eq:MPC_framework} that can be written as $\mathcal{U}^{i^\star}_k = \{u^{i^\star}_{p|k}\}_{p=0}^{N-1}$ and $\mathcal{X}^{i^\star}_k = \{x^{i^\star}_{p|k}\}_{p=0}^{N}$, and $x^{i^\star}_{N|k}\in\mathscr{X}^i_{\mathbb{T}_k}$. We are now constructing the following sequences for time $k+1$ using the optimal solutions $\mathcal{U}^{i^\star}_k$ and $\mathcal{X}^{i^\star}_k$:
        \begin{align}
            \mathcal{U}^i_{k+1} &= \{u^{i^\star}_{1|k},u^{i^\star}_{2|k},\cdots, \mu^i\},\label{eq:control_sequence_k1}\\
            \mathcal{X}^i_{k+1} &= \{x^{i^\star}_{1|k},x^{i^\star}_{2|k},\cdots,A_dx^{i^\star}_{N|k}+B_d\mu^i\}.\label{eq:state_sequence_k1}
        \end{align}
    where $\mu^i$ is closed-loop control law inside the terminal set $\mathscr{X}^i_{\mathbb{T}_k}$. Since $x^{i^\star}_{N|k}\in\mathscr{X}^i_{\mathbb{T}_k}$ and $\mathscr{X}^i_{\mathbb{T}_k}$ is a maximal control invariant set, there exists a closed-loop terminal control law $\mu^i\in \mathscr{U}^i$ which keeps the state $x^i_{N|k+1} = A_dx^{i^\star}_{N|k}+B_d\mu^i$ inside the terminal set. Therefore, $\mathcal{U}^i_{k+1}$ and $\mathcal{X}^i_{k+1}$ are the feasible solution of \eqref{eq:MPC_framework} (may not be the optimal solution). Further, since $x^i_{k+1} = x^{i^\star}_{1|k}$ and $u^i_{k+1} = u^{i^\star}_{1|k}$, therefore $(x^i_{k+1}, u^i_{k+1} )\in \mathscr{X}^i\times\mathscr{U}^i$. Similarly, we can show that the solution exists at $k+p,~~\forall p\in\mathbb{I}_2^\infty$. Hence, $\mathbb{OP}$ is recursively feasible.
\end{IEEEproof}

\subsection{Stability Analysis}
\label{subsec:stability}
In this subsection, we now demonstrate that the state trajectory $x^i_{p|k}$, generated by the implementation of control sequence $\mathcal{U}_k^i$ in accordance with the RHC methodology, converges to the reference state trajectory $x^i_{r}$.
\begin{lem}
    \label{lemma:lemma_2}
    Consider the cost function $J^i_k$ at time $k$ is a Lyapunov function. If the $\mathbb{OP}$ \eqref{eq:MPC_framework} is recursively feasible and 
        \begin{equation}
            \label{eq:convergence_condition}
            \begin{split}
                \min_{\substack{\mu^i\in\mathscr{U}^i,\\x^i_{N|k}\in\mathscr{X}^i_{\mathbb{T}_k}}}\Big(\Vert A_dx^i_{N|k}&+B_d\mu^i-x^i_{r}\Vert^2_P - \Vert x^i_{N|k}-x^i_{r}\Vert^2_P \\&+ \Vert x^i_{N|k}-x^i_{r}\Vert^2_Q + \Vert \mu^i\Vert^2_R\Big) \leq 0.
            \end{split}
        \end{equation}
    Then the closed-loop system is stable, that is, the following relation holds: $J^i_{k+1}\leq J^i_k$, for $k = 0, 1, 2, \ldots$
\end{lem}
\begin{IEEEproof}
    Given that the optimization problem $\mathbb{OP}$ is feasible at time $k$, we obtain the optimal control sequence $\mathcal{U}^{i^\star}_k$ and the associated sequence of state trajectories $\mathcal{X}^{i^\star}_k$, as explained in Lemma \hyperref[lemma:lemma_1]{1}. Therefore, an optimal control sequence $\mathcal{U}^{i^\star}_{k+1}$ is also attainable for the subsequent time $k+1$. Nevertheless, the sequence $\mathcal{U}^{i^\star}_{k+1}$ is unknown, we examine a sub-optimal control sequence $\mathcal{U}^i_{k+1}$ as delineated in \eqref{eq:control_sequence_k1} and the related state trajectories $\mathcal{X}^i_{k+1}$ as in \eqref{eq:state_sequence_k1}, with the associated cost function represented as $\widetilde{J}^i_{k+1}$. Given that $\mathcal{U}^i_{k+1}$ is sub-optimal, it follows that $\widetilde{J}^i_{k+1} \geq J^i_{k+1}$. the cost function $\widetilde{J}^i_{k+1}$ can be written as:
        \begin{equation}
            \label{eq:cost_func_subopt}
            \widetilde{J}_{k+1}^i = \Vert x_{N+1|k}^i-x_{r}^i\Vert^2_P + \sum_{p = 1}^{N}\Big\{\Vert x_{p|k}^i-x_{r}^i\Vert^2_Q + \Vert u_{p|k}^i\Vert^2_R\Big\}.
        \end{equation}
    Now consider terminal control law $u^i_{N|k} = \mu^i$. Therefore, using the reference of \eqref{eq:MPC_cost} it can be immediately shown that
        \begin{multline}
            \label{eq:cost_func_subopt_2}
            \widetilde{J}^i_{k+1} = J^i_{k} + \Vert A_dx^i_{N|k} + B_d\mu^i - x^i_{r}\Vert_P^2 + \Vert x^i_{N|k} - x^i_{r}\Vert_Q^2 \\~~~~~~~~~~+ \Vert \mu^i\Vert_R^2 -\Vert x^i_{N|k} - x^i_{r}\Vert_P^2 - \Big[\Vert x^i_{0|k} - x^i_{r}\Vert_Q^2 +\Vert u^i_{0|k} \Vert_R^2\Big].
        \end{multline}
    Since the term $\Big[\Vert x^i_{0|k} - x^i_{r}\Vert_Q^2 +\Vert u^i_{0|k} \Vert_R^2\Big] > 0$ and the condition \eqref{eq:convergence_condition} holds, therefore it follows from \eqref{eq:cost_func_subopt_2} that: $\widetilde{J}^i_{k+1} - J^i_{k} \leq 0$. Further, since $\widetilde{J}^i_{k+1} \geq J^i_{k+1}$, therefore $J^i_{k+1}\leq J^i_k$.
\end{IEEEproof}
\begin{rem}
    The condition \eqref{eq:convergence_condition} suggests that there exists at least one closed-loop control law $\mu^i$ which makes $\Big(\Vert A_dx^i_{N|k} +B_d\mu^i-x^i_{r}\Vert^2_P - \Vert x^i_{N|k}-x^i_{r}\Vert^2_P + \Vert x^i_{N|k}-x^i_{r}\Vert^2_Q + \Vert \mu^i\Vert^2_R\Big) \leq 0$. Further, the control invariance property of the set $\mathscr{X}^i_{\mathbb{T}_k}$ ensures the existence of such a terminal control law $\mu^i$.
\end{rem}

\subsubsection{Computation of terminal cost weight matrix \texorpdfstring{$P$}{P}}
\label{ssubsec:compute_P}
In the condition \eqref{eq:convergence_condition}, matrix $P$ is a design parameter. The terminal set $\mathscr{X}^i_{\mathbb{T}_k}$ is designed to be a forward invariant which ensures the recursive feasibility of the $\mathbb{OP}$ (Lemma \ref{lemma:lemma_1}). This implies that inside $\mathscr{X}^i_{\mathbb{T}_k}$, the constraints on the states and controls, i.e., $\mu_i\in\mathscr{U}^i$ and $x^i_{N|k}\in\mathscr{X}^i_{\mathbb{T}_k}$, will always be satisfied. Therefore, in order to compute $P$, the problem in \eqref{eq:convergence_condition} inside $\mathscr{X}^i_{\mathbb{T}_k}$ can be considered as follows:
    \begin{equation}
        \label{eq:convergence_optimization}
        \begin{split}
            \min ~~\Vert A_dx^i_{N|k}&+B_d\mu^i-x^i_{r}\Vert^2_P - \Vert x^i_{N|k}-x^i_{r}\Vert^2_P \\&+ \Vert x^i_{N|k}-x^i_{r}\Vert^2_Q + \Vert \mu^i\Vert^2_R
        \end{split}
    \end{equation}
The cost function in \eqref{eq:convergence_optimization} is
    \begin{equation*}
        \begin{split}
            \mathcal{L} &= ~\Vert A_dx^i_{N|k}+B_d\mu^i-x^i_{r}\Vert^2_P - \Vert x^i_{N|k}-x^i_{r}\Vert^2_P \\&~~+ \Vert x^i_{N|k}-x^i_{r}\Vert^2_Q + \Vert \mu^i\Vert^2_R,\\
            &= ~\Big(A_d(x^i_{N|k} - x^i_{r})+B_d\mu^i+(A_d - I_6)x^i_{r}\Big)^\top P\\&~~~~~\Big(A_d(x^i_{N|k} - x^i_{r})+B_d\mu^i+(A_d - I_6)x^i_{r}\Big)\\&~~~~-(x^i_{N|k}-x^i_{r})^\top P(x^i_{N|k}-x^i_{r})\\&~~~~+ (x^i_{N|k}-x^i_{r})^\top Q(x^i_{N|k}-x^i_{r}) + \mu^iR\mu^i,
        \end{split}
    \end{equation*}
    \begin{equation}
        \label{eq:lagrangian}
        \begin{split}
           ~~~~~~ &= ~(x^i_{N|k}-x^i_{r})^\top(A_d^\top PA_d-P+Q)(x^i_{N|k}-x^i_{r})\\&~ + (x^i_{N|k}-x^i_{r})^\top A_d^\top PB_d\mu^i + \mu^{i^\top}B_d^\top PA_d(x^i_{N|k}-x^i_{r})\\&+\mu^{i^\top}(R+B_d^\top PB_d)\mu^i +2\mu^{i^\top}B_d^\top P(A_d-I_6)x^i_{r}\\&~~+ 2(x^i_{N|k}-x^i_{r})^\top A_d^\top P(A_d-I_6)x^i_{r}.
        \end{split}
    \end{equation}
The equation \eqref{eq:lagrangian} can be rewritten as follows:
    \begin{equation}
        \label{eq:lagrangian_2}
        \mathcal{L} = \zeta^{i^\top} M \zeta^i + f^\top\zeta^i.
    \end{equation}
where $\zeta^i = \left[\begin{smallmatrix}x^i_{N|k}-x^i_{r}\\\mu^i\end{smallmatrix}\right]$, $M = \left[\begin{smallmatrix}A^\top_dPA_d-P+Q & A^\top_dPB_d\\B^\top_dPA_d & R+B^\top_dPB_d
\end{smallmatrix}\right]$, and $f = 2\left[\begin{smallmatrix}A_d^\top P(A_d-I_6)x^i_{r}\\B_d^\top P(A_d-I_6)x^i_{r}\end{smallmatrix}\right]$. Therefore, from \eqref{eq:lagrangian_2}, the optimal $\zeta^{i}$ we get, $\zeta^{i^\star} = -\frac{1}{2}M^{-1}f$. Hence, inside the terminal set,
    \begin{equation}
        \label{eq:lagrangian_min}
            \min~\mathcal{L} ~= \mathcal{L}(\zeta^{i^\star}) ~= -\frac{1}{4}f^\top M^{-1}f.
    \end{equation}
Consequently from \eqref{eq:lagrangian_min} it can be noted that the condition \eqref{eq:convergence_condition} inside the terminal set to be satisfied if $M$ is positive definite ($M \succ 0$) which yields the following matrix inequality:
    \begin{equation}
        \label{eq:matrix_inequality}
        A^\top_dPA_d-P+Q - A^\top_dPB_d\Big(R+B^\top_dPB_d\Big)^{-1}B^\top_dPA_d \succ 0.
    \end{equation}
The matrix $P$ can be computed by solving \eqref{eq:matrix_inequality} using an LMI solver, such as YALMIP \cite{J_Lofberg_2004}.

The overall implementation of the control scheme for CAV is summarized in Algorithm \hyperref[algo:algorithm_mpc]{5}.
    
    \begin{algorithm}
        \small{\caption{MPC algorithm for each AV}
        \textbf{Offline:} Compute $P$ by solving \eqref{eq:matrix_inequality}.\\
        \textbf{Online:}
        \begin{algorithmic}[1]
            \State Compute $v^i_{x_{r}}$ using Algorithm \hyperref[algo:algorithm_1]{1} for the signalized junction.
            \For {$k \geq 0$}
                \State Linearize the collision avoidance constraint \eqref{eq:safe_distance_const} using \Statex\hspace{1.5em}Algorithm \hyperref[algo:algorithm_linearapprox]{2}.
                
                \State Compute $\mathscr{X}^i_{\mathbb{T}_k}$ using Algorithm \hyperref[algo:algorithm_terminal]{3}.
                
                \State Run $\mathbb{OP}$ and generates a control sequence $\mathcal{U}^i_k = \mspace{-3mu}\{u^i_{p|k}\}_{p=0}^{N-1}$.
                
                \State Consider the first element $u^i_{0|k}$ of $\mathcal{U}^i_k$ and apply $u^i_{0|k}$ in \eqref{eq:ss_dist_dynamics} 
                \Statex\hspace{1.5em}with initial state $x^i_{0|k}$, and get $x^i_{1|k}$.
                
                \State Share the predicted state sequence $\{x^i_{p|k}\}_{p=0}^{N}$ of $i^{th}$ AV with
                \Statex\hspace{1.5em}the neighbouring AVs.
                
                \State If any AV considered as \textbf{LCV} finds enough space for lane
                \Statex\hspace{1.5em}change at the target lane, it changes its lane using Algorithm \Statex\hspace{1.5em}\hyperref[algo:algorithm_lanechange]{4}.
                
                \State $k \gets k+1$.
            \EndFor
        \end{algorithmic}
        \label{algo:algorithm_mpc}}
    \end{algorithm}
\textbf{Explanation for Algorithm \hyperref[algo:algorithm_mpc]{5}:} Initially, we calculated the terminal cost weight matrix $P$ offline using \eqref{eq:matrix_inequality}. The reference longitudinal velocity $v^i_{x_{r}}$ for each AV is computed in step $1$ using Algorithm \hyperref[algo:algorithm_1]{1} for the signalized junction.

\emph{Steps 3-7:} In step $3$, each AV linearized the collision avoidance constraint \eqref{eq:safe_distance_const} using the procedure mentioned in Algorithm \hyperref[algo:algorithm_linearapprox]{2}. Then, the maximal control invariant set ($\mathscr{X}^i_{\mathbb{T}_k}$) is constructed in step $4$ using Algorithm \hyperref[algo:algorithm_terminal]{3}. Each AV then runs the constrained optimization problem $\mathbb{OP}$ and generates an optimal control sequence $\mathcal{U}^i_k$ in step $5$. In step $6$, each AV apply the first control input $u^i_{0|k}$ from the sequence $\mathcal{U}^i_k$ with initial state $x^i_{0|k}$ and generates $x^i_{1|k}$. Then, they share their predicted state sequence $\{x^i_{p|k}\}_{p=0}^N$ with neighbouring AVs in step $7$.

\emph{Steps 8-9:} The step $8$ is significant when any AV at $k = k_{lc}$ chooses to change its lane. Subsequently, they change their lane using Algorithm \hyperref[algo:algorithm_lanechange]{4}. The time step is finally updated in step $9$.
\section{Numerical Example}
\label{sec:numerical_example}
We consider a CAV system consisting of $20$ AVs, which are moving on a $3$-lane road. The system matrix $A$ and the input matrix $B$ of the continuous-time system \eqref{eq:ss_cont_dynamics} are generated using the initial value of longitudinal velocity $v_x^i = 15$ m/s and lateral velocity $v^i_y = 0$ m/s. The other system parameters considered as follows: $m^i = 2050$ kg, $I^i_z = 3344$ $\text{kg}\cdot\text{m}^2$, $l^i_f = 1.1$ m, $l^i_r = 1.58$ m, $C^i_f = 65000$ N/rad, $C^i_r = 85000$ N/rad. The system we considered here is linear time-invariant. Therefore, the discretized dynamics of each AV with $t_s = 0.2$s as in \eqref{eq:ss_dist_dynamics}, where\\

$A_d = \begin{bmatrix}
        1&0&0.2&0&0&0\\
        0&1&0&0.0984&3&0.0936\\
        0&0&1&0&0&0\\
        0&0&0&0.079&0&-0.2149\\
        0&0&0&0.013&1&0.0703\\
        0&0&0&0.0493&0&0.0428
    \end{bmatrix}$,\\
    and $B_d = \begin{bmatrix}
        0.02&0&0.2&0&0&0\\
        0&1.0405&0&2.6421&0.5069&3.8306
    \end{bmatrix}^\top$.\\
    
We also considered that each AV entered the control zone at a different time step. The following is the sequence of the entrance time (in seconds) of each AV into the control zone: 

$T = \{0.0,~ 0.4,~ 1.4,~ 1.4,~ 2.6,~ 4.0,~ 4.2,~ 4.2,~ 6.0,~ 6.4,~ 6.4,~ \\~~~~~~~~~~~~7.4,~ 9.6,~ 10.8,~ 12.6,~ 14.0,~ 14.0,~ 16.2,~ 17.6,~ 18.6\}$. 

Based on the entry, each AV is assigned a number by the coordinator. The road width we considered here is $100$ m with approx $33.3$ m of lane width. The $y$-coordinates of the centerline trajectory of each lane are as follows: for lane-$1$: $\xi_y = 683.25$, for lane-$2$: $\xi_y = 649.95$, and for lane-$3$: $\xi_y = 616.65$. Vehicles entered in lane-$1$: \{AV$1$, AV$4$, AV$7$, AV$10$, AV$13$, AV$16$, AV$19$\}; vehicles entered in lane-$2$: \{AV$2$, AV$5$, AV$8$, AV$11$, AV$14$, AV$18$\}; vehicles entered in lane-$3$: \{AV$3$, AV$6$, AV$9$, AV$12$, AV$15$, AV$17$, AV$20$\}. Each vehicle entered the zone with the longitudinal velocity of $15$ m/s. The state constraints for $i^{th}$ AV are: $0\leq \xi^i_{x_k}\leq 1600$, $600\leq \xi^i_{y_k}\leq 700$, $0\leq v^i_{x_k}\leq 30$, $-5\leq v^i_{y_k}\leq 5$, $-0.1745\leq \psi^i_k\leq 0.1745$, $-1\leq \omega^i_k\leq 1$, and input constraints for $i^{th}$ AV are: $-8\leq a^i_{k}\leq 6$, $-0.75\leq \delta^i_k\leq 0.75$. The static gap is considered as $\gamma = 5$, and the prediction horizon is considered as $N = 20$. Then by choosing matrices $Q = \text{diag}(10^{-9}, 1, 10, 10, 1, 1)$ and $R = \text{diag}(5,5)$, we offline computed the matrix $$P = \left[\begin{smallmatrix}
 0.000465 & 0 & 0 & 0 & 0 & 0 \\
 0 & 0.4912 & -0.000023 & -0.1583 & 0.1558 & -0.0431\\
 0 & -0.000023 & 4.4834 & 0.000129 & -0.000469 & -0.000029\\
 0 & -0.1583 & 0.000129 & 1.9709 & -0.9814 & -0.6115\\
 0 & 0.1558 & -0.000469 & -0.9814 & 4.3164 & 0.1713\\
 0 & -0.0431 & -0.000029 & -0.6115 & 0.1713 & 0.4723\end{smallmatrix}\right]$$

 The Multi-parametric toolbox \cite{M_Herceg_2013} is used for the computation of all the polyhedral sets (precursor set, terminal set), and the optimization problem is solved using the MATLAB function \emph{fmincon} in MATLAB R$2020$a in an Intel Core i$5$ $1.20$ GHz processor, $8$ GB RAM, $64$-bit operating system. It is observed that each AV takes on average $3.653$ seconds for each iteration, which includes online terminal set computation and $\mathbb{OP}$ to run, where for computing online terminal set, each AV takes on average $2.981$ seconds.
 
\begin{figure}[thbp]
        \centering
        \includegraphics[height = 4.5cm, width = 9.1cm]{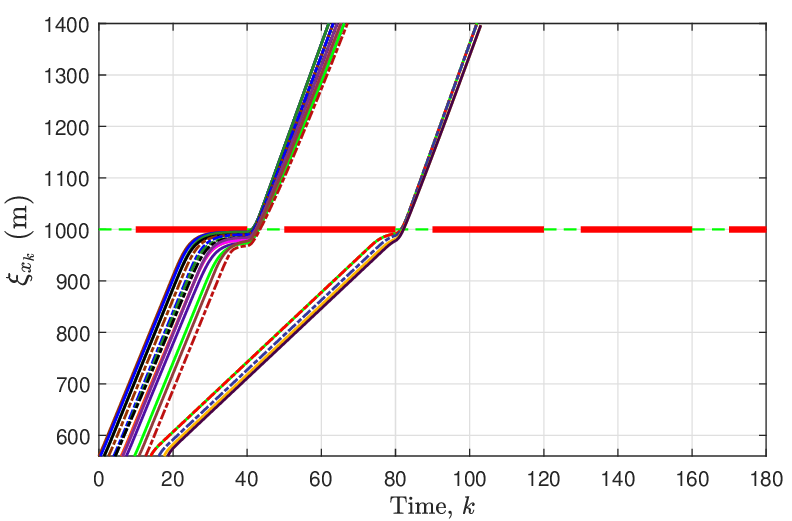}
        \caption{Longitudinal Positions of all AVs ($20$ AVs) crossing $SJ$. Horizontal red lines (solid) denote red signal duration, and horizontal green lines (dashed) denote green signal duration.}
        \label{fig:longtudinal_position}
    \end{figure}

\begin{figure}[thbp]
        \centering
        \includegraphics[height = 4.8cm, width = 9.1cm]{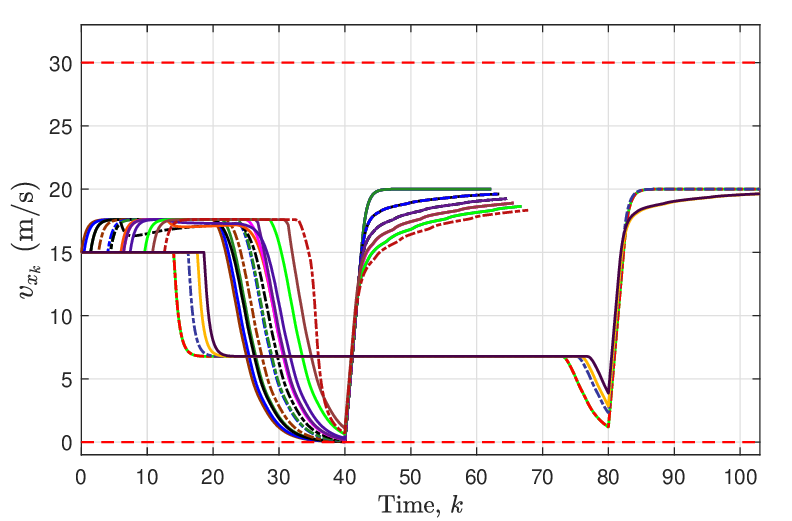}
        \caption{Longitudinal Velocity of all AVs ($20$ AVs). Horizontal red lines (dashed) denote maximum and minimum velocity bounds.}
        \label{fig:longtudinal_velocity}
    \end{figure}

\begin{figure}[thbp]
        \centering
        \includegraphics[height = 4.8cm, width = 9.1cm]{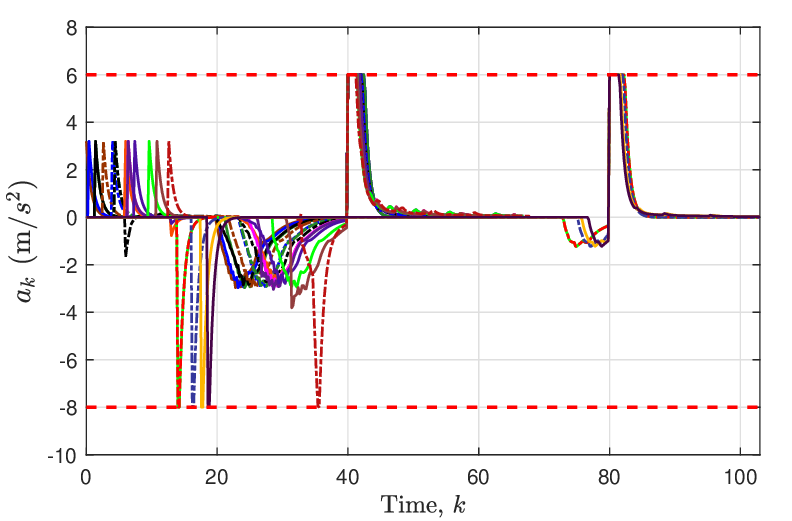}
        \caption{Longitudinal Acceleration of all AVs ($20$ AVs). Horizontal red lines (dashed) denote maximum and minimum acceleration bounds.}
        \label{fig:longtudinal_acceleration}
    \end{figure}

 The longitudinal position of all the AVs is depicted in Fig.~(\ref{fig:longtudinal_position}). The \emph{critical density} near the junction point is considered here to be $15$. Therefore, only $15$ AVs can stop at the upcoming red signal phase and pass through the $SJ$ at the next green signal, whereas other $5$ AVs have to slow down so that they can cross the $SJ$ at the later green signal phase. It can be observed that due to the imposition of the terminal constraint, the AVs have adjusted their speed profiles and slowed down accordingly to stop at the red signal while maintaining their minimum safe distance. Further, Fig.~(\ref{fig:longtudinal_velocity}) and Fig.~(\ref{fig:longtudinal_acceleration}) show that the constraints on longitudinal velocity and acceleration for each AV are also satisfied. \\
    
\begin{figure*}[thpb]
        \centering
        \includegraphics[height = 8.5cm, width = 18.3cm]{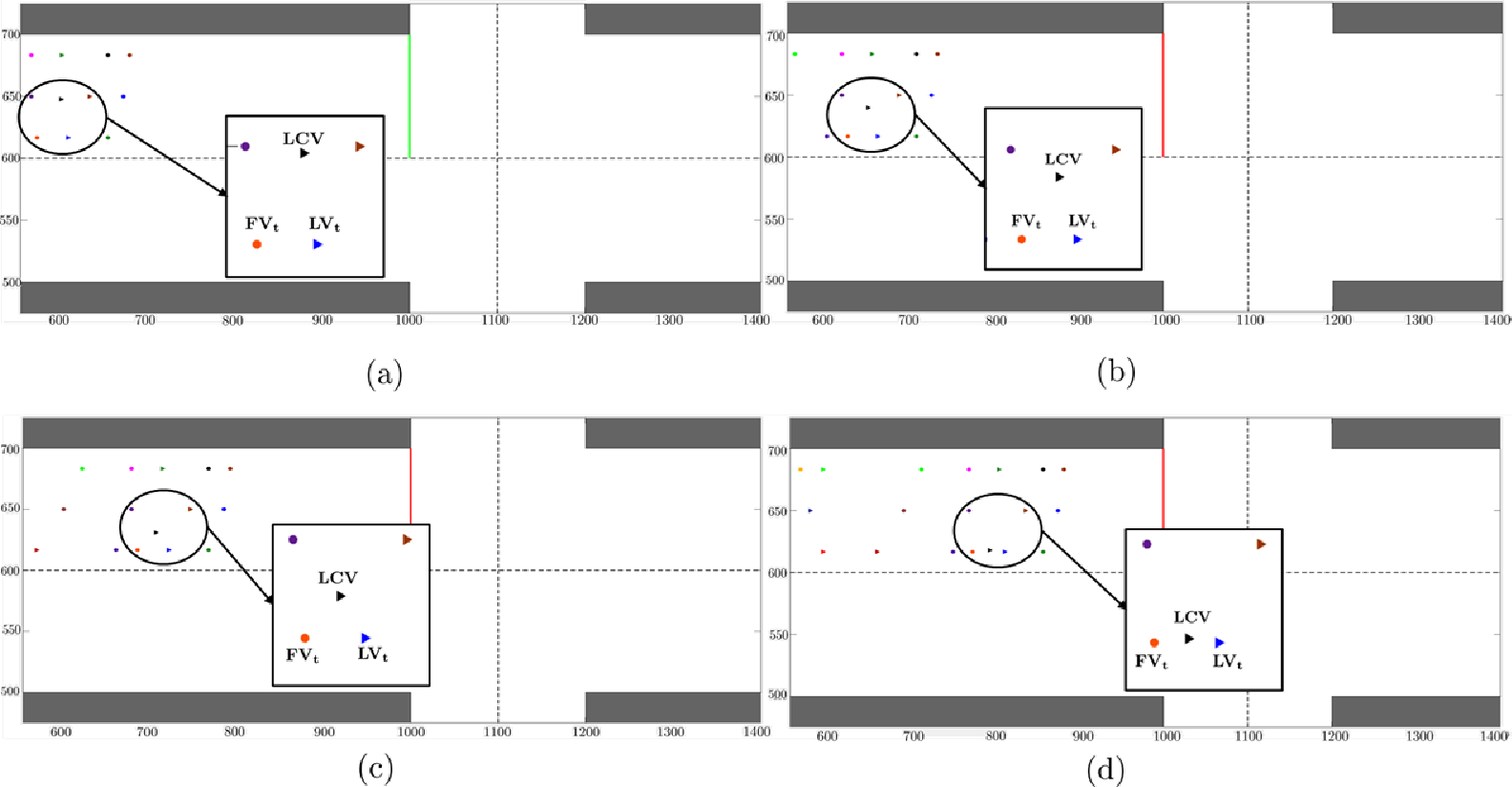}
        \caption{Snapshots of lane change of $\mathbf{LCV}$ from lane-$2$ to lane-$3$ in between $\mathbf{LV_t}$ and $\mathbf{FV_t}$. The road junction is shown in white colour. The solid green and red lines indicate green and red traffic signals, respectively. The dashed black lines separate the two sides of the road. Each side consists of three lanes.}
        \label{fig:lane_change_snapshots}
    \end{figure*}

    \emph{Lane change scenario:}~ In lane change scenario, we considered AV$8$, that is, $\mathbf{LCV}$ changes its lane from lane-$2$ to lane-$3$. Fig.~(\ref{fig:lane_change_snapshots}) depict the lane change scenario where $\mathbf{LCV}$ changes its lane to lane-$3$ in between $\mathbf{LV_t}$ and $\mathbf{FV_t}$ safely.
    
\begin{figure}[thpb]
        \centering
        \includegraphics[height = 6.8cm, width = 9.1cm]{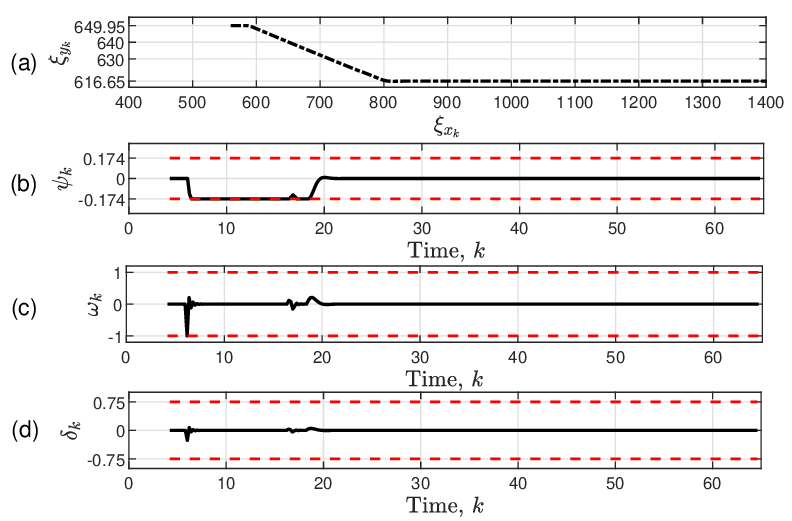}
        \caption{(a) Positions trajectory of $\mathbf{LCV}$, (b) Yaw angle of $\mathbf{LCV}$, (c) Yaw rate of $\mathbf{LCV}$, (d) Steering angle of $\mathbf{LCV}$.}
        \label{fig:lateral_position_yaw_angV_steer_HV}
    \end{figure}

\begin{figure}[thpb]
        \centering
        \includegraphics[height = 5.4cm, width = 9.1cm]{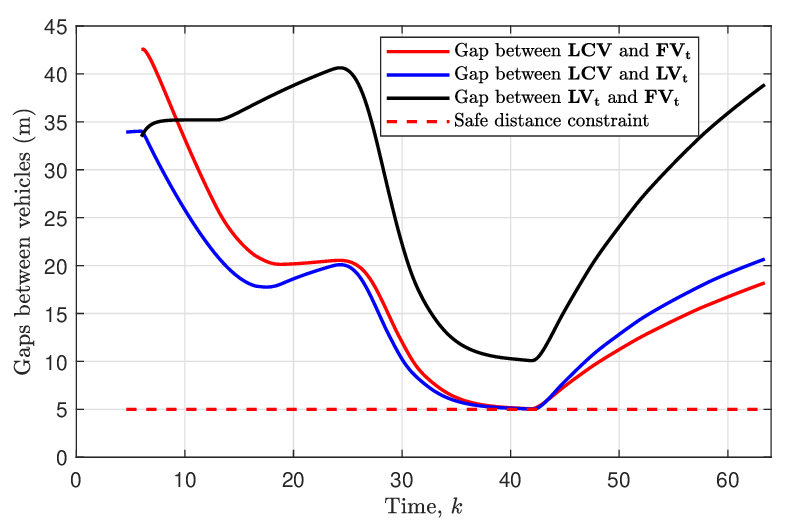}
        \caption{Minimum safe distance constraint (Red dashed line), gaps between $\mathbf{LCV}$ and $\mathbf{FV_t}$ (Red solid line), gaps between $\mathbf{LCV}$ and $\mathbf{LV_t}$ (Blue line), gaps between $\mathbf{LV_t}$ and $\mathbf{FV_t}$ (Black Line).}
        \label{fig:gaps_HV_FVt_LVt}
    \end{figure}

    Fig.~(\ref{fig:lateral_position_yaw_angV_steer_HV}) demonstrates the position trajectory, yaw angle, yaw rate, and steering angle of $\mathbf{LCV}$, which also satisfied the constraints. Also the minimum safe distance is maintained among the vehicles $\mathbf{LCV}$, $\mathbf{LV_t}$, and $\mathbf{FV_t}$ during lane change scenario is shown in Fig.~(\ref{fig:gaps_HV_FVt_LVt}). The MATLAB simulation video of the proposed algorithm is available at \url{https://github.com/rudrasen10/Coordinated-control-of-AVs}.


\section{Conclusion}
This study extends the dual-mode MPC theory presented in \cite{R_Sen_2024} to multi-input systems. This work proposes a combined strategy for congestion mitigation and cooperative lane manoeuvring that develops upon existing MPC-based frameworks. By incorporating the terminal constraint, we guarantee stability and feasibility. As CAVs become a more prevalent component of future transportation systems, this integrated perspective may facilitate more efficient and adaptable traffic management. Future efforts will focus on robust MPC to handle the external disturbances and the uncertainties due to the intervention of human-driven vehicles in traffic. Moreover, subsequent research will aim to incorporate the design of data-driven Model Predictive Control (MPC) while addressing the inadequate knowledge of the system model of AVs.  




\bibliographystyle{IEEEtran}
\bibliography{biblio}
\end{document}